\documentclass{article}

\begin{document}

\title{Realism in the Realized Popper's Experiment}

\author{Geoffrey Hunter}{address={
Department of Chemistry,
York University, Toronto, Canada  M3J 1P3.\ \ ghunter@yorku.ca}}

\begin{abstract}
The realization of Karl Popper's EPR-like experiment by
Shih and Kim (published 1999) produced the result that Popper hoped for: no
``action at a distance'' on one photon of an entangled pair when a
measurement is made on the other photon.  This experimental result
is interpretable in local realistic terms: each photon has a definite
position and transverse momentum most of the time; the position measurement on
one photon (localization within a slit) disturbs the transverse momentum of that
photon in a non-predictable way in accordance with the uncertainty
principle; however, there is no effect on the other photon (the photon that is not in a slit)
no action at a distance.  The position measurement (localization within a slit) of the one
photon destroys the coherence (entanglement) between the
photons;~i.e.~decoherence occurs.

This realistic (albeit retrodictive) interpretation of the Shih-Kim realization of what Popper called his
``crucial experiment'' is in
accord with Bohr's original concept of the nature of the uncertainty
principle, as being an inevitable effect of the disturbance of the
measured system by the measuring apparatus.  In this experiment
the impact parameter of an incident photon with the centerline of the slit
is an uncontrollable parameter of each individual photon scattering event;
this impact parameter is variable
for every incident photon, the variations being a statistical aspect of the beam of
photons produced by the experimental arrangement.

These experimental results are also in accord with the
proposition of Einstein, Podolski and Rosen's 1935 paper: that quantum mechanics provides only a
statistical, physically incomplete, theory of microscopic physical processes, for the quantum mechanical
description of the experiment does not describe or explain the individual photon scattering events
that are actually observed; the angle by which
an individual photon is scattered is not predictable, because the photon's impact parameter
with the centerline of the slit is not observable, and because the electromagnetic interaction between
the photon and the matter forming the walls of the slit is not calculable.
\end{abstract}

\maketitle

\section{Popper's Concept of the Crucial Experiment}
Karl Popper was a philosopher who was deeply concerned with the interpretation of quantum
mechanics since its inception in 1925.   In a book originally published in 1956
he proposed an experiment which he described as {``an extension of the Einstein-Podolsky-Rosen
argument''} \cite[pp.ix,27-30]{Popper}.  In summary: two photons are emitted simultaneously from
a source that is mid-way
between (and colinear with) two slits A and B; the coincident diffraction pattern is observed beyond
each slit twice: once with both slits present, and again with one slit (slit B)
``wide open''.\footnote{T.~Angelidis has restated this as ``slit B being absent''.}

Popper (in common with Einstein, Podolsky and Rosen and with the proponents of the Copenhagen
interpretation of quantum mechanics) believed that {\em quantum theory predicts} that localization of
a photon within slit A will not only cause it to diffract in accordance with Heisenberg's uncertainty
relation, but will also cause the other photon at the location of slit B to diffract by the
same angle - regardless of whether slit B is present or not.
Such spatially separated correlations would violate the causality principle of Special Relativity; i.e.~that
physical interactions cannot travel faster than the speed of light; thus Einstein called them
``spooky actions at a distance''.

Popper was ``inclined to predict'' \cite[p.29]{Popper} that in an actual experimental test, the photon at
the location of the absent slit B would not diffract (``scatter'') when its partner photon (with which it
is in an entangled state) is localized within slit A.  He emphasized that ``this does not mean that quantum
mechanics (say, Schr\"odinger's formalism) is undermined''; only that ``Heisenberg's claim is undermined
that his formul\ae\ are applicable to all kinds of indirect measurements''.

Popper \cite[p.62]{Popper} notes that Heisenberg agreed that {\em retrodictive} values of the position
{\em and} the momentum can be known by knowing the position of a particle (e.g.~as it passes through
a small slit) followed by a measurement of the momentum of the particle after it has passed through
the slit - from the position in the detection plane where it is located.   Feynman has also emphasized that
the uncertainty principle does not exclude  {\em retrodictive} inferences about the simultaneous
position and momentum of a particle \cite[Ch.2, pp.2-3]{Feynman}, but  the uncertainty
principle does exclude precise {\em prediction of both} position or momentum.

Popper also notes
 that momentum values are usually inferred from two sequential position measurements; in this regard
he concurs with  Heisenberg, who wrote \cite[p.62]{Popper}:
\begin{quote}
``The \ldots most fundamental method of measuring velocity [or momentum] depends on the
determination of position at two different times \ldots it is possible to determine {\em with any
desired degree of accuracy} what the velocity [or momentum] of the particle {\em was} before the
second measurement was made''
\end{quote}
Indeed precise
inferences of particle positions and momenta are widely used in the analysis of  observations in
high-energy particle accelerators, by which the modern plethora of ``fundamental particles'' have been
discovered.  This is emphasized by another quote from Popper  \cite[p.39]{Popper}:
\begin{quote}
``my assertion [is] that most physicists who honestly believe in the \\\ \ Copenhagen interpretation
do not pay any attention to it in actual practice''.
\end{quote}

\section{The Real Experiment of Shih and Kim}

	Yoon-Ho Kim and Yanhua Shih carried out a modern realization of Popper's experiment;
their report was published three times
\cite{Shih1,Shih2,Shih3}.  Other experiments that don't confirm the
widely held interpretation of quantum mechanics have also
been reported \cite{Bell1972}.

Kim and Shih note that Popper's thought experiment was designed to show:
\begin{quote}
``that a particle can have both precise position \\ and momentum at the same time'' --
\cite[p.1849, 2$^{\rm nd}$ paragraph]{Shih1}, and that
\end{quote}
\begin{quote}
``it is astonishing to see that the experimental \\ results agree with Popper's prediction'' --
 \cite[p.1850, 2$^{\rm nd}$ paragraph]{Shih1}.
\end{quote}

\subsubsection{Experimental Parameters}\label{exptdata}
\begin{itemize}
\item the source is a CW (continuous wave) argon-ion laser producing highly
monochromatic ultra-violet radiation of wavelength, $\lambda= 351.1$ nm;
\item a pair of entangled visible photons ($\lambda= 702.2$ nm)
are produced by Spontaneous Parametric
Down-Conversion (SPDC) in a BBO ($\beta$ barium borate) crystal;
\item the width of slit A and of slit B (when present) $= 0.16$ mm;
\item the diameter of the avalanche photodiode detectors (D1 and D2) $= 0.18$ mm;
\item during the measurements, D1 was in a fixed position (\underline{not} scanned along the $y$-axis)
close to a collection lens (of 25 mm focal length) with the lens close to slit A; this was
designed to direct {\em every} photon passing through slit A into detector D1; the photons thus detected were used
simply to trigger the coincidence circuit to look for a photon with detector D2 (500 mm beyond slit B).
\end{itemize}

\subsection{Transverse Momentum: Slit B Present}\label{diffHUP}

The classical (wave-optics) theory of diffraction \cite[pp.214-216]{Longhurst} based upon Huygens
principle\footnote{that each point in a beam is a source, the resulting beam being the result of interference
between the radiation from all the sources}
predicts that the diffraction pattern produced by a beam of light incident upon a single slit in a screen perpendicular
to the incident direction of the beam, will have a first minimum intensity at:
\begin{eqnarray}\label{classical}
 y  = \frac{\lambda\,D}{s}
\end{eqnarray}
where $y$ is the transverse displacement of the diffraction minimum from the incident direction of the beam
in the plane of the detecting screen (at distance $D$ from the slit of width $s$);
$\lambda$ is the wavelength of the light and $y$ is naturally proportional to $D$.

Feynman has shown \cite{Feynman} that the classical formula (\ref{classical}) can be re-interpreted as an effect of the
Heisenberg Uncertainty Relation, for substitution of the de Broglie relation, $\lambda=h/P$ between a photon's
wavelength, $\lambda$ and its momentum, $P$, into eqn.(\ref{classical}) produces:
\begin{eqnarray}\label{HUPdiffdelta}
 y\, P  = \frac{h\,D}{s}  \quad\quad  {\rm or}  \quad\quad  \Delta y\, \Delta P_y  = h
\end{eqnarray}
because the photon's {\em transverse} momentum, $\Delta P_y$, is $P\times y/D$.
Thus diffraction towards the first diffraction minimum is interpreted as being the
measure of the ``uncertainty'' in the transverse (i.e.~in the $y$ direction) momentum of the photon, $\Delta P_y$,
caused by location of its $y$-coordinate within the width of the slit, $\Delta y=s$.

I have carried out calculations showing that (\ref{classical}) and (\ref{HUPdiffdelta}) are in accord
with the first diffraction minimum in the experimental curve in Figure 5 of \cite{Shih1} when slit B is present.

{It is noteworthy that Kim and Shih report} \cite[top of p.1856]{Shih1}:
\begin{quote}
``the {\em single} detector counting rate of D2 is basically the same as \\ that of the coincidence
counts except for a higher counting rate''.
\end{quote}
In other words: the outer (wider) diffraction peak (the curve for slit B present) of Figure 5 of
\cite{Shih1} is obtained regardless of whether the coincidence detection circuit is active or not.
This observation indicates that this peak is produced by the diffraction of the photons incident
upon slit B {\em regardless} of their entanglement with the photons incident upon slit A;
i.e~it is a single-photon phenomenon; indeed it is nothing more than the central diffraction maximum
predicted by the classical theory of light \cite[p.214]{Longhurst}.

The higher counting rate is explained by the effective size of the source only being about
0.16 mm diameter for coincidence counting, whereas for non-coincidence counting it is the full diameter
of the laser beam of 3 mm.

\subsection{Transverse Momentum: Slit B Absent}

The inner curve of Figure 5 of \cite{Shih1} is drawn from the experimental coincidence counts obtained
when slit B is ``wide open''.  This shows that the photons are deflected
up to 0.9 mm, and that from 0.9 mm to 1.45 mm the detection rate is constant and close to
zero.\footnote{These observed counting rates are only zero within the plotting precision of Figure 5
of \cite{Shih1}; the observed counting rates were not recorded as actual, recorded values in any
of the published records of the experiments.}
This range (0.9 to 1.45 mm) over which the counting rate is constant precludes discernment
of a diffraction minimum (unlike the curve for slit B present,
which has one point above zero as the onset of the
second diffraction maximum); there are in fact 5 data points
($y= 1.0 - 1.45$ mm) all of which have the same (very small) value; this suggests that the
origin of this peak is {\em not} diffraction through a slit.

It is noteworthy that Kim and Shih report \cite[p.1856,$2^{\rm nd} paragraph$]{Shih1}:
\begin{quote}
``the single detector counting rate of D2 keeps \\\ \ \ constant in the entire scanning range''
\end{quote}
In other words: when the coincidence circuit is switched off  D2 detects the same count rate at
all values of $y$ at which it was placed ($y=0$ to $y=1.45$ mm); this would be
a horizontal straight line if added to Figure 5 of \cite{Shih1}.
This measurement was simply seeing the beam emanating from the source towards D2
with a uniform intensity over the scanning range of $\approx$ 3 mm ($y=\pm 1.5$ mm).

Kim and Shih also report \cite[p.1856,$2^{\rm nd} paragraph$]{Shih1}:
\begin{quote}
``the width of the pattern is found to be much narrower \\\ \ \ than the actual size of the diverging SPDC
beam at D2''.
\end{quote}
The probable cause of this narrow peak is a convolution of the finite size of the
source,\footnote{a cylinder 3 mm diameter and 3 mm long} with the geometry of possible
coincidences; if it were interpreted as a diffraction pattern, then in terms of eqn.(\ref{HUPdiffdelta}),
there would be an observed violation of the Heisenberg Uncertainty Principle by a factor of about 3.

\section{A Retrodictive Realist Account}
\begin{itemize}
\item A pair of photons produced by SPDC is in an entangled state from the moment of generation
 until one of them enters a slit; entanglement means that their positions
and momenta are correlated; knowledge of the position of one photon allows one to infer the
position of the other photon; likewise for their momenta.
\item When one photon enters a slit it interacts with that slit and this destroys the coherence
(entanglement) between them; i.e.~decoherence occurs.
The interaction can be attributed to the photon being a
localized electromagnetic wave \cite{HunterWad},\footnote{In \cite{HunterWad} the photon is an ellipsoidal
soliton of length $\lambda$, and diameter $\lambda/\pi$.}
which interacts with the electrons in the surface of the solid
that forms the slit.  That photons are localized waves is supported experimentally by the production
of laser pulses as short (in time) as two optical periods \cite{Corkum};
thus the photon cannot be longer than two wavelengths along its direction of propagation.
\item Measurement of the diffracted position ($y$ coordinate) of a photon
(coincidence detection by D2) with slit B absent, allows one to calculate not only the momentum
vector of this photon as it travels from the source to D2, but also the momentum of the other
photon as it travels from the source to slit A; however when this latter photon enters slit A its
interaction with the walls of the slit causes it to diffract at an angle which is predictable only
statistically - in accord with the uncertainty principle.

\item Thus coincidence measurements \underline{with slit B absent} provide the positions of both photons
(from the detection of a photon having passed through slit A) with a precision equal to the
width of slit A.  Likewise the measurement of the deflected position ($y$ coordinate) of a photon by D2
allows one to calculate the momentum vectors of {\em both} photons of the entangled pair -- during their
trajectories from the source to the plane of slit A (for one photon), and from the source
to the scanning plane of D2 for the other photon.  These {\em in principle}, precise, retrodictive
calculations of the trajectories of both photons are unfortunately limited in precision by the actual
experimental results because of the relatively large size of the non-point source.\footnote{a cylinder
3 mm diameter and 3 mm long}

\item It is especially noteworthy that \underline{individual events}\footnote{the generation and detection
of a particular entangled photon pair} are not limited by the uncertainty principle: any diffraction of
a photon to a position of D2 smaller than the $y$-coordinate of the first diffraction minimum will
yield a position-momentum product that is smaller than Planck's constant -- even when slit B
is present.  In particular, the most probable diffraction angle (to the top of the central peak)
yields a transverse momentum of zero, which when multiplied by the uncertainty in its
position (the slit width of 0.16 mm) yields an uncertainty product of zero !
\end{itemize}

\subsection*{Concluding Remarks}
Karl Popper regarded his experiment as a ``crucial'' test of the inconsistency between
the inferred non-locality of quantum mechanics and the causality principle of Special Relativity.
While Shih and Kim conceded \cite[p.1858,last paragraph]{Shih1} that:
``Popper and EPR were correct in the prediction of the physical outcomes of their experiments.'',
their subsequent sentences (in the same paragraph),
(``\ldots Popper and EPR made the same error \ldots'') are incongruous; Popper and EPR made no error - they
agreed with Bohr, Heisenberg and other proponents of the Copenhagen interpretation \cite{Cushing}
that quantum theory apparently predicts an instantaneous action at a distance on one particle of an
entangled pair, when a measurement is made on the other particle of the pair;
Popper and EPR's crucial point was that if such actions at a distance are
not in fact observed (as in the Shih-Kim experiment), then quantum theory must be an incomplete (only
statistical) theory of the physical world, and as a statistical theory that does not describe individual
events, it is entirely consistent with the causality principle of Special Relativity.

The above interpretation of the Shih-Kim experiment in locally realistic terms,
is not easily extended to the interpretation of some other experiments: Bohm's EPR
experiment involving the Bell inequalities \cite[Ch.11]{Holland}, and
 the various single-particle, double-path experiments conducted with
particles as large as $C_{60}$ molecules \cite{Zeilinger}.

\section*{acknowledgments}
Thomas Angelidis brought the Shih-Kim experiment to the
author's attention at conferences in Baltimore (1999) and Berkeley (2000), and
Yanhua Shih provided helpful elaborations of the published accounts of the experimental work.
The Natural Sciences and Engineering Research Council of Canada provided financial support.

\end{document}